\documentclass[a4paper,conference,final]{IEEEtran}
\usepackage{siunitx}
\DeclareSIUnit \dBm {dBm}
\DeclareSIUnit \dB {dB}
\DeclareSIUnit \Mbps {Mbps}
\DeclareSIUnit \Gbps {Gbps}
\DeclareSIUnit \mph {mph}

\usepackage{graphicx}
\usepackage{pstricks,pst-node,pst-tree}
\usepackage{amsmath}
\usepackage{multirow}
\usepackage{url}
\usepackage{multicol}
\usepackage[printonlyused,withpage]{acronym}
\usepackage{amsmath, amssymb}
\usepackage{bm}
\usepackage{epstopdf}
\usepackage{amsfonts}
\usepackage{amsmath}

\usepackage{algorithm}
\usepackage[noend]{algpseudocode}
\makeatletter
\newcommand*{\algrule}[1][\algorithmicindent]{\makebox[#1][l]{\hspace*{.5em}\vrule height .75\baselineskip depth .35\baselineskip}}%

\newcount\ALG@printindent@tempcnta
\def\ALG@printindent{
    \ifnum \theALG@nested>0
        \ifx\ALG@text\ALG@x@notext
            \addvspace{-0.5pt}
        \else
            \unskip
            \ALG@printindent@tempcnta=1
            \loop
                \algrule[\csname ALG@ind@\the\ALG@printindent@tempcnta\endcsname]
                \advance \ALG@printindent@tempcnta 1
            \ifnum \ALG@printindent@tempcnta<\numexpr\theALG@nested+1\relax
            \repeat
        \fi
    \fi
    }
\usepackage{etoolbox}
\patchcmd{\ALG@doentity}{\noindent\hskip\ALG@tlm}{\ALG@printindent}{}{\errmessage{failed to patch}}
\makeatother

\newtheorem{remark}{Remark}[section]

\title{Optimized Certificate Revocation List Distribution for Secure V2X Communications}
\author{\IEEEauthorblockN{Giovanni Rigazzi\IEEEauthorrefmark{1}, Andrea Tassi\IEEEauthorrefmark{1}, Robert J. Piechocki\IEEEauthorrefmark{1}, Theo Tryfonas\IEEEauthorrefmark{3}, Andrew Nix\IEEEauthorrefmark{1}} 
\IEEEauthorblockA{\IEEEauthorrefmark{1}Department of Electric and Electronic Engineering, University of Bristol, UK} 
\IEEEauthorblockA{\IEEEauthorrefmark{3}Department of Civil Engineering, University of Bristol, UK\\ E-mail: \{g.rigazzi, a.tassi, r.j.piechocki, theo.tryfonas,  andy.nix\}@bristol.ac.uk}}

\begin{document}
\maketitle

\begin{abstract}
The successful deployment of safe and trustworthy Connected and Autonomous Vehicles (CAVs) will highly depend on the ability to devise robust and effective security solutions to resist sophisticated cyber attacks and patch up critical vulnerabilities. Pseudonym Public Key Infrastructure (PPKI) is a promising approach to secure vehicular networks as well as ensure data and location privacy, concealing the vehicles' real identities. Nevertheless, pseudonym distribution and management affect PPKI scalability due to the significant number of digital certificates required by a single vehicle. In this paper, we focus on the certificate revocation process and propose a versatile and low-complexity framework to facilitate the distribution of the Certificate Revocation Lists (CRL) issued by the Certification Authority (CA). 
CRL compression is achieved through optimized Bloom filters, which guarantee a considerable overhead reduction with a configurable rate of false positives. Our results show that the distribution of compressed CRLs can significantly enhance the system scalability without increasing the complexity of the revocation process. 
\end{abstract}



\section{Introduction}

Connected and Autonomous Vehicles (CAVs) rely heavily upon a wide spectrum of heterogeneous technologies combining autonomous driving and vehicle-to-everything (V2X) communications, with the goal of achieving social and economic benefits, such as enhanced road safety, reduced traffic congestion and air pollution~\cite{AtkinsCAV}. 
Besides innovative emergency services and infotainment applications leveraging Global Positioning Systems (GPS) and cellular systems, Dedicated Short Range Communications (DSRC) connectivity allows vehicles to exchange real-time information provided by on-board sensor devices, and make decisions based on multiple factors, including road conditions and traffic status~\cite{7355569}. 

One of the most critical issues concerning the deployment of such vehicular networks is how to efficiently integrate cyber security mechanisms, and ensure trustworthiness and anonymity of exchanged data. Due to the their intrinsic characteristics, V2X communications can be targeted by numerous cyber attacks and security threats, ranging from injection of bogus information or node impersonation, to malicious location tracking and privacy leakage~\cite{petit2015potential}. As a result, security for CAVs has been subject of intensive joint research activities among automotive industry, standardization and regulatory bodies, public authorities and academia, resulting in a plethora of research projects and initiatives across the world~\cite{zhao2013challenges}. 

As part of the IEEE 1609 Wireless Access in Vehicular Environments (WAVE) suite, IEEE 1609.2 represents the reference standard for security and privacy adopting a Public Key Infrastructure (PKI), where vehicle authentication is achieved through a Certification Authority (CA) or Trusted Third Party (TPP), in charge of issuing legitimate digital certificates and binding vehicle identity to its public key~\cite{IEEE16092}. 
An authenticated vehicle can then use its corresponding private key to digitally sign each outgoing packet, whereas sender identity verification is performed by using the public key included in the certificate assigned by the CA. 
Moreover, the CA is responsible for revoking certificates associated with corrupted or misbehaving entities. To this end, IEEE 1609.2 defines Certificate Revocation Lists (CRL) containing the identities of the revoked certificates, which are periodically updated and disseminated by the CA in the vehicular network~\cite{raya2006certificate}. Upon the reception of a new message, a vehicle can identify a legitimate sender by verifying whether the corresponding certificate is not published in the CRL.

To further preserve data privacy and limit vehicle traceability, vehicle identities (VIDs) can be replaced with multiple abstract short-lived identifiers, i.e., pseudonyms, thus realizing a Pseudonym PKI (PPKI)~\cite{petit2015pseudonym}. The pseudonym credentials are issued by the CA, which is also responsible for verifying the eligibility of a vehicle to exchange data by storing its VID. Therefore, location privacy is preserved, as two consecutive messages are signed under two distinct and unlikable pseudonyms. However, this comes at the price of a significant increase in the CRL size, which in turn undermines the scalability and efficiency of the revocation process~\cite{7054434}. 

In this paper, we propose a low-complexity framework for ensuring trustworthy communications, aiming to reduce the overhead of the CRL distribution via optimized Bloom filter compression. 
Authors in~\cite{raya2006certificate} first adopt Bloom filters to compress the CRLs and reduce the amount of data disseminated. The resulting Compressed Certificate Revocation Lists (C$^2$RL) are then broadcast, while certificate validation is quickly performed by checking the Bloom filter associated with the latest C$^2$RL update. Similarly,~\cite{raya2007eviction} proposes the Revocation using C$^2$RL (RC$^2$RL) protocol, where the dissemination of compressed lists is achieved through Road Side Units (RSU) and mobile units. In addition, a quantitative analysis of the protocol performance is presented, showing the existing tradeoff between computational complexity and probability of false positives. Bloom filter compression is also employed in~\cite{haas2011efficient}, where vehicles locally compress and store the list of revoked certificates by generating the revocation keys with the help of optimized CRLs, which are disseminated in a V2V epidemic fashion. However, these solutions do not specify optimal values of the parameters associated with Bloom filters and do not consider the case of multiple certificates assigned to a single vehicle. 
Differently from the proposed approaches, we aim to evaluate the efficiency of C$^2$RLs in PPKI vehicular networks, and introduce an optimization framework to jointly minimize the filter size and the number of hash functions employed, according to a predefined probability of false positives. Our results show that the CRL distribution process can benefit from the significant overhead reduction, which can be characterized through the compression gain, without causing an increase in complexity. We also demonstrate the C$^2$RLs effectiveness in comparison with a standard CRL approach in an urban setting through large-scale network simulations. 

The rest of the paper is organized as follows. Section~\ref{Sec:SystemModel} describes the network model adopted as well as the certificate compression and the CRL distribution. The proposed optimization framework is illustrated in Section~\ref{Sec:OptModel}. In Section~\ref{Sec:NumRes}, we analyze the numerical results obtained, whereas the final conclusions are drawn in Section~\ref{Sec:Conclusions}.

\section{System Model}\label{Sec:SystemModel}
In this section, we describe the vehicular network and the main entities involved in our system model. We also discuss the fundamental aspects of the C$^2$RL issuance and distribution as well as the procedure to compress the certificates and perform the verification. 

\subsection{Network Model}
\begin{figure}[t]
    \begin{center}
        \includegraphics[width=0.8\columnwidth]{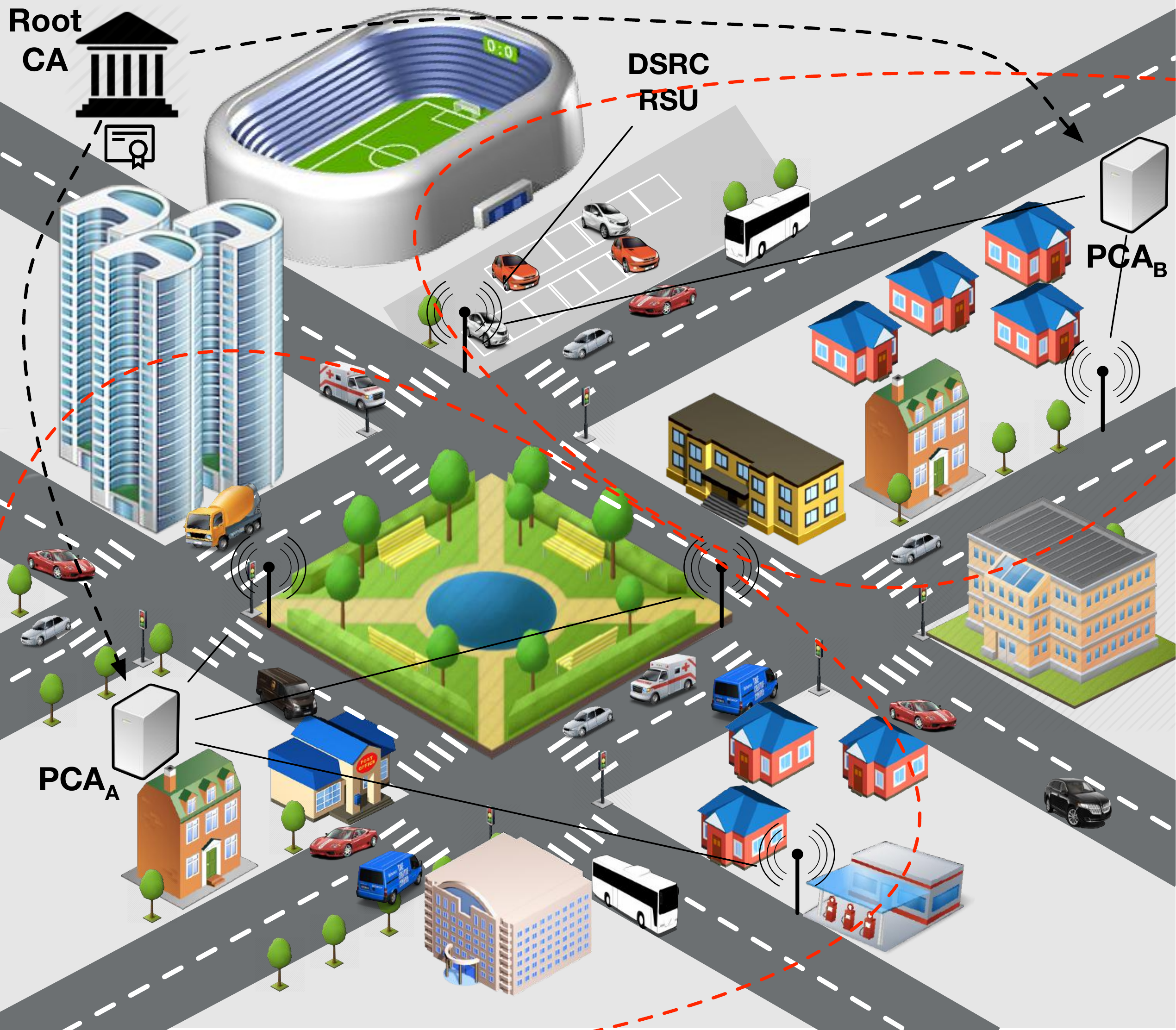}
        \caption{Illustration of the system model, where two PCAs are connected to a single RCA. }
        \label{fig:Topology}
    \end{center}
    \vspace{-12pt}
\end{figure}

In a typical hierarchical PPKI setup, a Root CA (RCA) coordinates the CAV authentication within a predefined jurisdictional area, such as a city, region, county, etc., by registering vehicles and assigning long-term certificates.  Fig.~\ref{fig:Topology} shows the reference scenario considered in this paper. A certain number of Pseudonym CAs (PCA) are also connected to the RCA through wired links, and are responsible for issuing pseudonyms and CRLs. 
We assume that RCA and PCAs are equipped with sufficient resources in terms of storage and computation, and cannot be compromised by potential attackers. Moreover, the RCA maintains the mapping of short-term credentials to the long-term identity of the vehicles. 
A number of Road-Side Units (RSU) are deployed along the roads, each connected to a single PCA via a wired backhaul network, while V2I wireless connectivity is achieved by employing DSRC interfaces, such as IEEE 802.11p or ETSI ITS-G5. To mitigate the potential lack of DSRC due to the RSU sparse deployment, we also assume that CAVs are supplied with a cellular radio interface, e.g., 3GPP LTE-A. Although cellular systems introduce additional delay and present limited applicability for safety critical applications, we expect that the integration of these two technologies will represent a key feature of next-generation V2X communications~\cite{7513432}. CAVs sign and broadcast safety-related messages, i.e., Cooperative Awareness Messages (CAM) and Decentralized Environmental Notification Messages (DENM), attaching the sender certificate, and are provided with a tamper-resistant Hardware Security Module (HSM) storing the cryptographic material. 

\subsection{Attacker Model}
We assume that an internal adversary is able to inject bogus information in terms of fake messages through a legitimate private/public key pair and a related pseudonym certificate. An attacker may also perform a Denial-of-Service (DoS) attack by broadcasting messages with the goal of reducing network resources necessary to reliably exchange safety messages. We assume that anomaly detection algorithms are capable of identifying these threats and triggering the eviction process.

\subsection{Certificate Compression}\label{subsec.CC}
\begin{figure}[t]
    \begin{center}
\includegraphics[width=0.6\columnwidth]{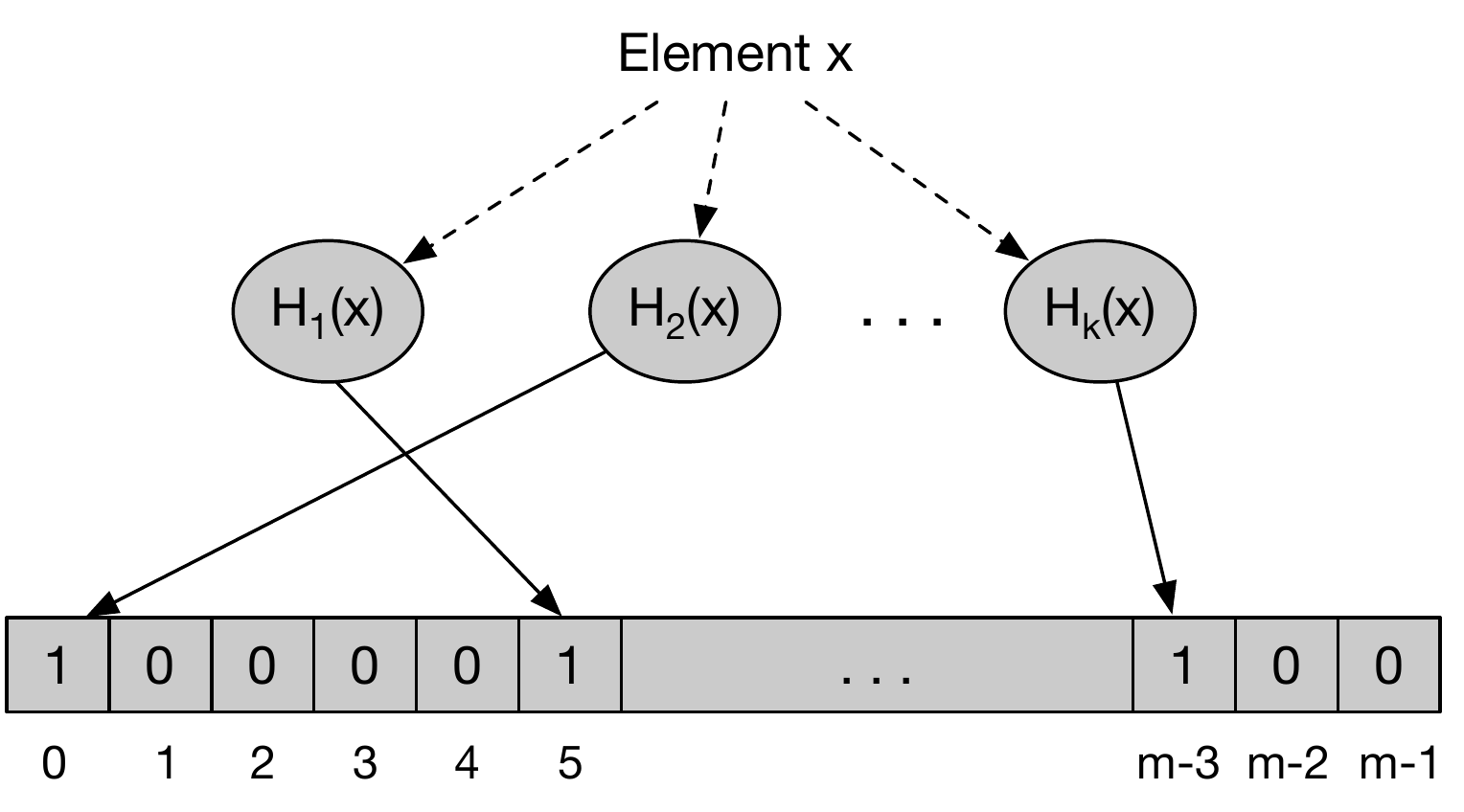}
        \caption{Example of Bloom filter of size $m$ bits and $k$ hash functions.}
        \label{fig:Bloom}
    \end{center}
    \vspace{-20pt}
\end{figure}

Following the PPKI approach~\cite{khodaei2015key}, we assume that each vehicle is assigned a certain number of short-term certificates containing different pseudonyms and pairs of public/private keys. 
In passive revocation schemes~\cite{ma2008pseudonym}, the pseudonym certificate lifetime is minimized, thus vehicles and PCAs need to frequently communicate to initiate the certificate renewal or pseudonym refill procedure. This requirement is usually dictated by the need for minimizing the vulnerability window, i.e., the time between a vehicle is declared illegitimate and subsequent pseudonym refill requests are denied, and all the associated pseudonyms are expired.  By contrast, our approach seeks to limit the frequency of pseudonym refills according to the pseudonym change strategy adopted and the storage capability\footnote{The problem of how to efficiently change the pseudonyms and determine the change rate is out of the scope of this paper.}. 
Efficient compression of CRLs can be accomplished with the help of Bloom filters. 
A Bloom filter is a probabilistic data structure typically adopted to verify whether a certain element belongs to a set~\cite{tarkoma2012theory}. 
Such a filter is characterized by a probability of false positives, i.e., probability that an element not included in the set is detected due to multiple hash collisions, while false negatives cannot occur. As shown in Fig.~\ref{fig:Bloom}, a Bloom filter corresponds to a sequence of $m$ bits set to 0. To add an element $x$, a set of $k$ independent hash functions $H_1, \ldots, H_k$ are employed. The output of each hash function matches one of the filter elements and sets to 1 the corresponding bit, while an element previously set to 1 cannot be altered. 
The verification procedure is then performed by checking the bits corresponding to the output of the hash functions. If all the corresponding bits are set to 1, an element is assumed to be contained in the filter with a certain probability, whereas negative outcomes are always true if at least one of the filter bits is 0. 
For a target number of filter elements $m$, the probability of false positives $\delta$ is given by~\cite{tarkoma2012theory}:
\begin{equation}
\delta(m,k) \doteq \left[1-\left(1-\frac{1}{m}\right)^{kn}\right]^k,\label{eq.deltaBloom}
\end{equation}
being $n$ the number of elements to add into the filter.
Hence, the filter accuracy is influenced by the size, the number of employed hash functions and $n$, as the larger the set of elements, the higher the probability of obtaining a false positive. Moreover, Bloom filters involve a small amount of computational overhead for insertion and search operations~\cite{haas2011efficient}. 


\subsection{CRL Issuance and Distribution}\label{Sec:CRL_description}

\begin{figure}[t]
    \begin{center}
        \includegraphics[width=0.95\columnwidth]{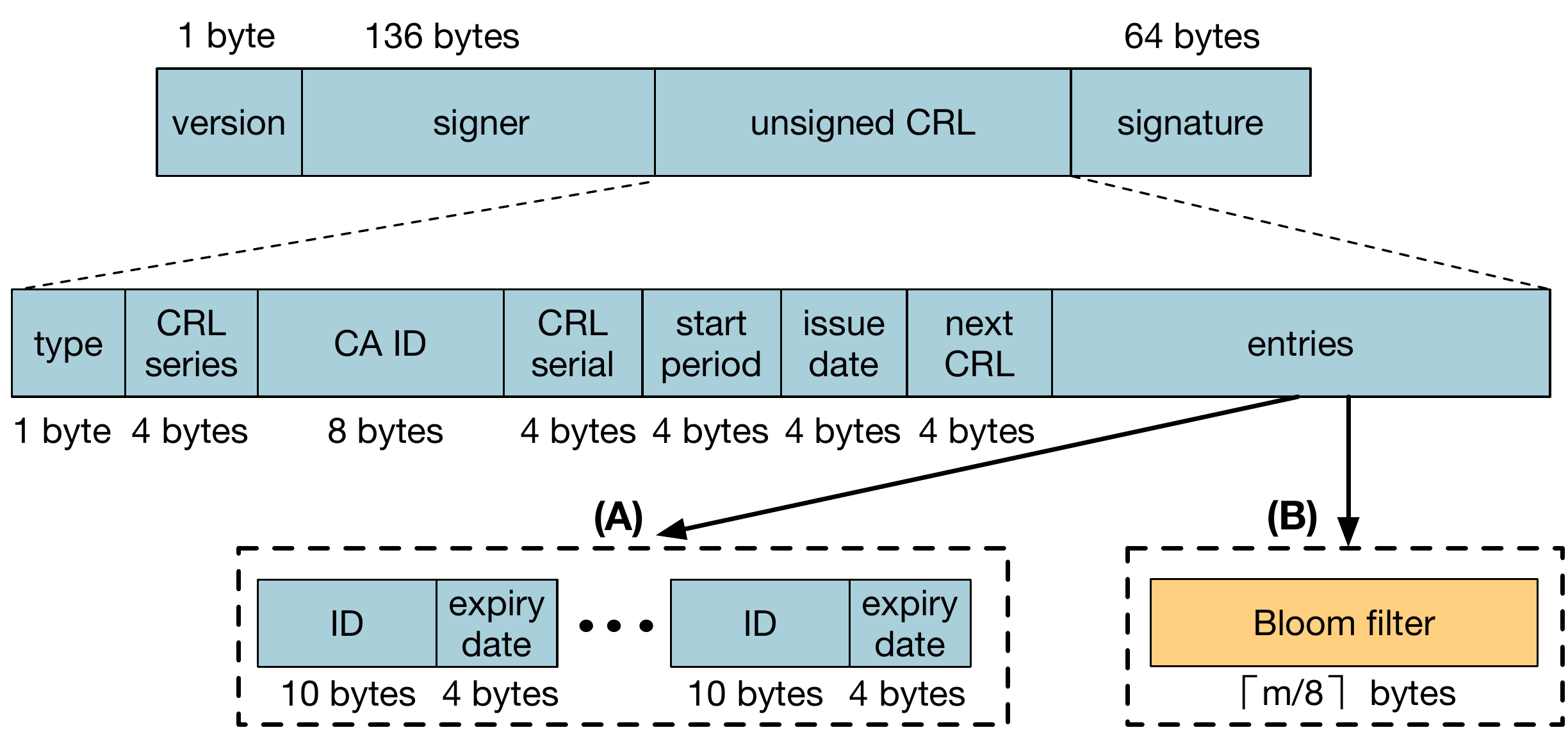}
        \caption{Structure of a WAVE CRL~\cite{IEEE16092}: standard CRL (A) and C$^2$RL (B).}
        \label{fig:CRL_structure}
    \end{center}
    \vspace{-20pt}
\end{figure}

Every time a misbehaving vehicle is identified, a new CRL needs to be issued and sent to the registered CAVs. As a result, all non-expired certificates associated with the evicted vehicle must be revoked, otherwise any pseudonym still valid may be used to sign the outgoing traffic\footnote{This assumption entails linkability among pseudonyms, which limits data privacy. The analysis of the tradeoff between overhead and privacy is left for future work.}. We also assume that a revocation authority, e.g., a government agency, is in charge of recognizing malicious vehicles and informing the RCA. 
As shown in Fig.~\ref{fig:CRL_structure}, a WAVE CRL consists of \emph{(i)} a header, including a \emph{version} field set to 1, a \emph{signer} field containing information on the CA issuing and signing the CRL, and a \emph{signature} field carrying the signature of the signer, and \emph{(ii)} the unsigned CRL field. A detailed description of each sub-field is reported in~\cite{IEEE16092}. 
To discuss the differences between a standard CRL (see Fig.~\ref{fig:CRL_structure}(A)) and a Bloom filter compressed CRL (see Fig.~\ref{fig:CRL_structure}(B)), we focus on the \emph{entries} sub-field, containing the identifiers of each revoked certificate. Specifically, in a standard WAVE CRL, each certificate is identified with an \emph{ID} field and an optional \emph{expiry date} field, used to enhance the efficiency of the certificate storage. By contrast, in our approach, a single Bloom filter of fixed size $m$ bits is carried by the \emph{entries} field. As a consequence, the size of this field remains constant as the number of revoked certificates increases, thus resulting in a significant reduction of the size of the CRLs distributed to the CAVs, as illustrated in Sec.~\ref{Sec:NumRes}. 
%
We assume that a C$^2$RL is issued by the RCA and delivered to its connected PCAs. 
Next, the RSUs receive the C$^2$RL forwarded by the PCAs and validate the attached signature. The C$^2$RL is then signed by the RSU and broadcast to the CAVs through DSRC connectivity, which verify the authenticity and store it in the HSM. 
Hence, untrustworthy vehicles can be quickly identified by verifying whether the certificates attached in the messages are contained in the filter transmitted in the latest C$^2$RL. 
We also point out that only the parameter $k$ needs to be notified to the CAVs in order to fulfill this process, as long as each CAV adopts the same implementation for the $k$ uniformly distributed hashing functions employed. To this end, we consider a single hashing function and provide it with different seed values\footnote{In our implementation, the $i$-th hashing function is simply obtained by setting the corresponding seed value to $i$.}. 

Being the generation of false positives inevitable, how to handle vehicle identities wrongly revoked represents an open research question. To overcome this issue, we adopt the approach proposed in~\cite{haas2011efficient} and provide the vehicles with a set of \emph{backup pseudonyms} to replace those pseudonyms generating false positives. Each vehicle periodically establishes whether the pseudonym in use triggers a false positive. In other words, each vehicle validates the pseudonym in use against the latest CRL update. Should a legitimate vehicle trigger a false positive, the pseudonym in use is replaced with one of the backup pseudonyms. This entails the provision of additional certificates to compensate for the number of pseudonyms discarded because of false positives. It is worth noting that the resulting false positive probability decreases exponentially as the number of backup pseudonyms per-vehicle increases~\cite{haas2011efficient}.

\section{Proposed Optimization Framework}\label{Sec:OptModel}
Let $\mathcal{S}=\{ k_1, ..., k_n\}$ be the set of certificates to be added into the Bloom filter, being $|\mathcal{S}| = n$ the cardinality of $\mathcal{S}$. We formulate our filter optimization (FO) model as follows\footnote{By $\mathbb{N}$ we denote the set of natural numbers.}:
\vspace{-2mm}\begin{align}
	\text{(FO)} &  \quad  \min_{m,k} \,\,  m \label{FO.of}\\
    \text{subject to} &   \quad \delta(m,k) \leq \hat{\delta}\label{FO.c1}\\
                      &   \quad k \geq 1, \quad m \geq 1 \label{FO.c2}\\
                      &   \quad m \in \mathbb{N}, \quad k \in \mathbb{N}.\label{FO.c3}
\end{align}
We observe that the FO model carries out a joint optimization of $m$ and $k$, having as objective the minimization of $m$, as per~\eqref{FO.of}. Constraint~\eqref{FO.c1} ensures that the probability of false positives is smaller than or equal to a target value $\Hat{\delta}$. Obviously, in order for a Bloom filter to exist, it should be at least one bit long and be associated with at least one hash function. 
These existing constraints are summarized in~\eqref{FO.c2}. Finally, constraint~\eqref{FO.c3} imposes $m$ and $k$ to be integer values.

In the attempt of solving the FO model, we relax constraint~\eqref{FO.c3}. In addition, we regard the function \mbox{$\tilde{\delta}(\tilde{m},\tilde{k}):\mathbb{R}^+ \times \mathbb{R}^+ \rightarrow [0,1]$} as the real expansion of $\delta(m,k)$ over the set of positive real values $\mathbb{R}^+$. As such, the relaxed FO  (rFO) model can be expressed as follows:
\begin{equation}
\text{(rFO)} \,\, \displaystyle\mathop{\arg\min}_{\tilde{m},\tilde{k}} \Big\{\tilde{m} \,\,\Big|\,\, \tilde{\delta}(\tilde{m},\tilde{k}) \leq \Hat{\delta} \land k \geq 1 \land m \geq 1  \Big\}.\label{eq.rFO}
\end{equation}

\begin{figure}[t]
    \begin{center}
        \includegraphics[width=1\columnwidth]{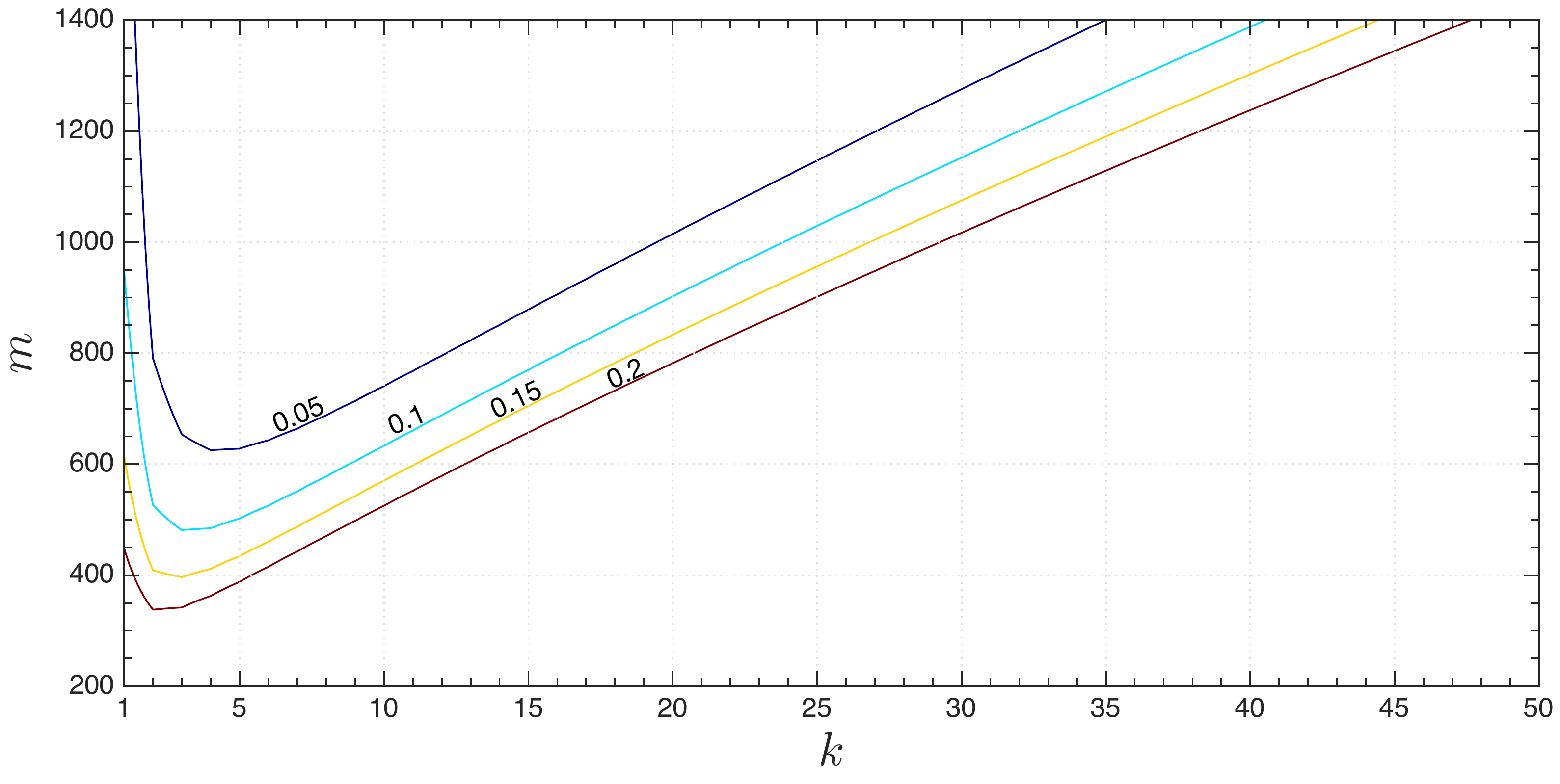}
        \vspace*{-7mm}
        \caption{Contour plot of $\delta(m,k)$, for $n = 10^2$. We only reported the level curves corresponding to the pairs $(m,k)$ where $\delta(m,k)$ is equal to $0.05, 0.1, 0.15$ and $0.2$.}
        \label{fig:Opt}
    \end{center}
        \vspace{-10pt}
\end{figure}

\begin{remark}\label{rem.1}
From~\eqref{eq.deltaBloom}, it follows that $\delta(m,k)$ goes to $0$ as $m$ tends to infinity, which is not surprising (see Section~\ref{subsec.CC}). Furthermore, from~\eqref{eq.deltaBloom}, we observe that as $m$ increases, $\delta(m,k)$ cannot increase. Hence, considering constraint~\eqref{FO.c1}, the more $\delta(m,k)$ approaches $\Hat{\delta}$, the more $m$ is likely to decrease, as also shown in Fig.~\ref{fig:Opt}. Obviously, the same observation also applies to $\tilde{\delta}(\tilde{m},\tilde{k})$.
\end{remark}

For a given value of $\Hat{\delta}$, we define the function $\tilde{m}(\tilde{k})$ providing the value of $\tilde{m}$ such that relation $\tilde{\delta}(\tilde{m}(\tilde{k}),\tilde{k}) = \Hat{\delta}$ holds. From~\eqref{eq.deltaBloom}, $\tilde{m}(\tilde{k})$ can be defined as follows:
\begin{equation}
\tilde{m}(\tilde{k}) \doteq \left[ 1 - \left( 1 - \Hat{\delta}^{-\frac{1}{\tilde{k}}} \right)^{\frac{1}{\tilde{k} n}} \right]^{-1}\label{eq.m}.
\end{equation}
We denote by $(\tilde{m}^*,\tilde{k}^*)$ the optimum solution of rFO. From Remark~\ref{rem.1}, we observe that the optimum solution of rFO shall satisfy condition $\tilde{\delta}(\tilde{m}^*,\tilde{k}^*) = \Hat{\delta}$. As such, the value of $\tilde{k}^*$ can be defined as follows:
\begin{equation}
\tilde{k}^* = \displaystyle\mathop{\arg\min}_{\tilde{k}} \Big\{\tilde{m}(\tilde{k})\Big\},
\end{equation}
while $\tilde{m}^*$ is simply equal to $\tilde{m}(\tilde{k}^*)$.

From~\eqref{eq.m}, we derive the first order derivative of $\tilde{m}(\tilde{k})$, which is~\cite{jeffrey2007table}:
\begin{equation}
\frac{\partial \tilde{m}}{\partial \tilde{k}} = \frac{T^\frac{1}{\tilde{k} n}}{(1-T^\frac{1}{\tilde{k} n})^2} \left[ \frac{ \Hat{\delta}^\frac{1}{\tilde{k}} \log(\Hat{\delta}) }{\tilde{k}^3 n T} - \frac{\log(T)}{\tilde{k}^2 n} \right],
\end{equation}
where
\begin{equation}
T \doteq 1 - \Hat{\delta}^\frac{1}{\tilde{k}}.
\end{equation}
We observe that the equation $\frac{\partial \tilde{m}}{\partial \tilde{k}} = 0$ has at least a real root iff the equation
\begin{equation}
\Hat{\delta}^\frac{1}{\tilde{k}} \log(\Hat{\delta}) - \tilde{k} T \log(T) = 0 \label{eq.solSeed}
\end{equation}
has at least a real root, as well. By resorting to the bisection strategy~\cite{moore1979methods}, it is numerically simple to observe that this circumstance occurs for practical values of $\tilde{m}$, $\tilde{k}$ and $\Hat{\delta}$\footnote{Specifically, we refer to $\tilde{m} \in [1, 2^{32}]$, $\tilde{k} \in [1, 10^3]$ and \mbox{$\Hat{\delta} \in [10^{-4}, 2 \cdot 10^{-1}]$}.}. Finally, we observe that the real root of~\eqref{eq.solSeed}, if it exists, is equal to $\tilde{k}^*$.

\setlength{\textfloatsep}{5pt}
\begin{algorithm}[t]
\floatname{algorithm}{Procedure}
\caption{Solution of FO}
\label{Alg.P1}
\begin{algorithmic}[1]
\State $\tilde{k}^* \gets \text{the real root of~\eqref{eq.solSeed}}$\label{sInit.1}
\State $\tilde{m}^* \gets \tilde{m}(\tilde{k}^*)$\label{sInit.2}
\State $i \gets 1$
\For{$k \gets \lfloor \tilde{k}^* \rfloor, \lceil \tilde{k}^* \rceil$}\label{mainSolLoopStart}
	\State $\mathbf{k}(i) \gets k$
	\State $\mathbf{m}(i) \gets \lfloor \tilde{m}^* \rfloor$
    \vspace{-0.3mm}\While {$\delta(\mathbf{m}(i),\mathbf{k}(i)) > \Hat{\delta}$}\label{wStart}
    	\State $\mathbf{m}(i) \gets \mathbf{m}(i) + 1$
    \EndWhile\label{wEnd}
    \vspace{-0.9mm}\State $i \gets i + 1$
\EndFor\label{mainSolLoopEnd}
    \For{$i \gets 1,2$}\label{SolTest1Start}
    	\vspace{-0.3mm}\If{$\delta(\mathbf{m}(i),\mathbf{k}(i)) > \Hat{\delta}$}
			\State $\mathbf{m}(i) \gets NaN$
			\State $\mathbf{k}(i) \gets NaN$
		\EndIf
    \EndFor\label{SolTest1End}
    
    \If{$\mathbf{m}(1) \neq \mathbf{m}(2)$}\label{SolTest2Start}
		\State $j \gets \text{\textit{index of the smallest element in} $\mathbf{m}$}$
    \Else
    	\State $j \gets \text{\textit{index of the smallest element in} $\mathbf{k}$}$
	\EndIf\label{SolTest2End}
\State{\Return $(\mathbf{m}(j),\mathbf{k}(j))$}
\end{algorithmic}
\end{algorithm} 

From the optimum solution $(\tilde{m}^*,\tilde{k}^*)$ of rFO, we derive the optimum solution  $(m^*,k^*)$ of FO, as per Procedure~\ref{Alg.P1}. In particular, for each value in vector $\mathbf{k} = [\lfloor \tilde{k}^* \rfloor, \lceil \tilde{k}^* \rceil]$, the for-loop at lines~\ref{mainSolLoopStart}-\ref{mainSolLoopEnd} and~\ref{SolTest1Start}-\ref{SolTest1End} (of Procedure~\ref{Alg.P1}) derives the minimum filter length that meets constraint~\eqref{FO.c1}. These values are then stored in vector $\mathbf{m}$. Lines~\ref{SolTest2Start}-\ref{SolTest2End} allow the procedure to select the solution associated with the smallest filter size or, if both the solutions refer to the same filter size, the procedure returns the one with the smallest number of hash functions. 




\section{Performance Evaluation}\label{Sec:NumRes}

In this section, we discuss the performance in terms of overhead reduction obtained by employing our optimized framework. To this end, we denote by $\mathcal{G}$ the compression gain, corresponding to the ratio between the size of a standard CRL and the size of a C$^2$RL adopting the Bloom filter compression. As described in Sec.~\ref{Sec:CRL_description}, a standard CRL consists of a fixed section of $230$ bytes and an additional $14$ bytes per each revoked certificate, whereas the size of a CCRL is equal to $230 + \lceil m/8 \rceil$ bytes. 
Finally, we consider a large-scale urban scenario and evaluate the $C^2RL$ gains through network simulations.

\begin{table}[t]
\renewcommand{\arraystretch}{1.3}
\centering
    \caption{System parameters considered in the Holloway scenario.}
{\scriptsize\begin{tabular}{|p{2cm}|c||p{2cm}|c|}
\hline
Parameter & Value & Parameter & Value \\ \hline \hline
Area size                    & \SI{3}{\kilo\meter} $\times$ \SI{2}{\kilo\meter} & Max. UDP packet size         & 1024 bits \\
Sim. duration                & \SI{3600}{\second} & Carrier freq. & \SI{5.89}{\giga\hertz} \\
SUMO through traffic factor~\cite{behrisch2011sumo}  & 7 & TX power          & \SI{20}{\milli\watt} \\
SUMO traffic count~\cite{behrisch2011sumo}           & 15 & Phy. layer bitrate & \SI{11}{\Mbps} \\
No. pseudonyms per-vehicle   & 1000 & Sensitivity             & \SI{-89}{\dBm} \\
CRL TX interval              & \SI{300}{\second} & Thermal noise     & \SI{-110}{\dB} \\
Optimized Bloom filter              & $\Hat{\delta} = 10^{-3}$, $n = 300$ &  & \\ \hline
	\end{tabular}}
    \label{tab:system_parameters}
\end{table}

\subsection{Optimal Values of $k$ and $m$}
\begin{figure}[t]
    \begin{center}
        \includegraphics[width=0.9\columnwidth]{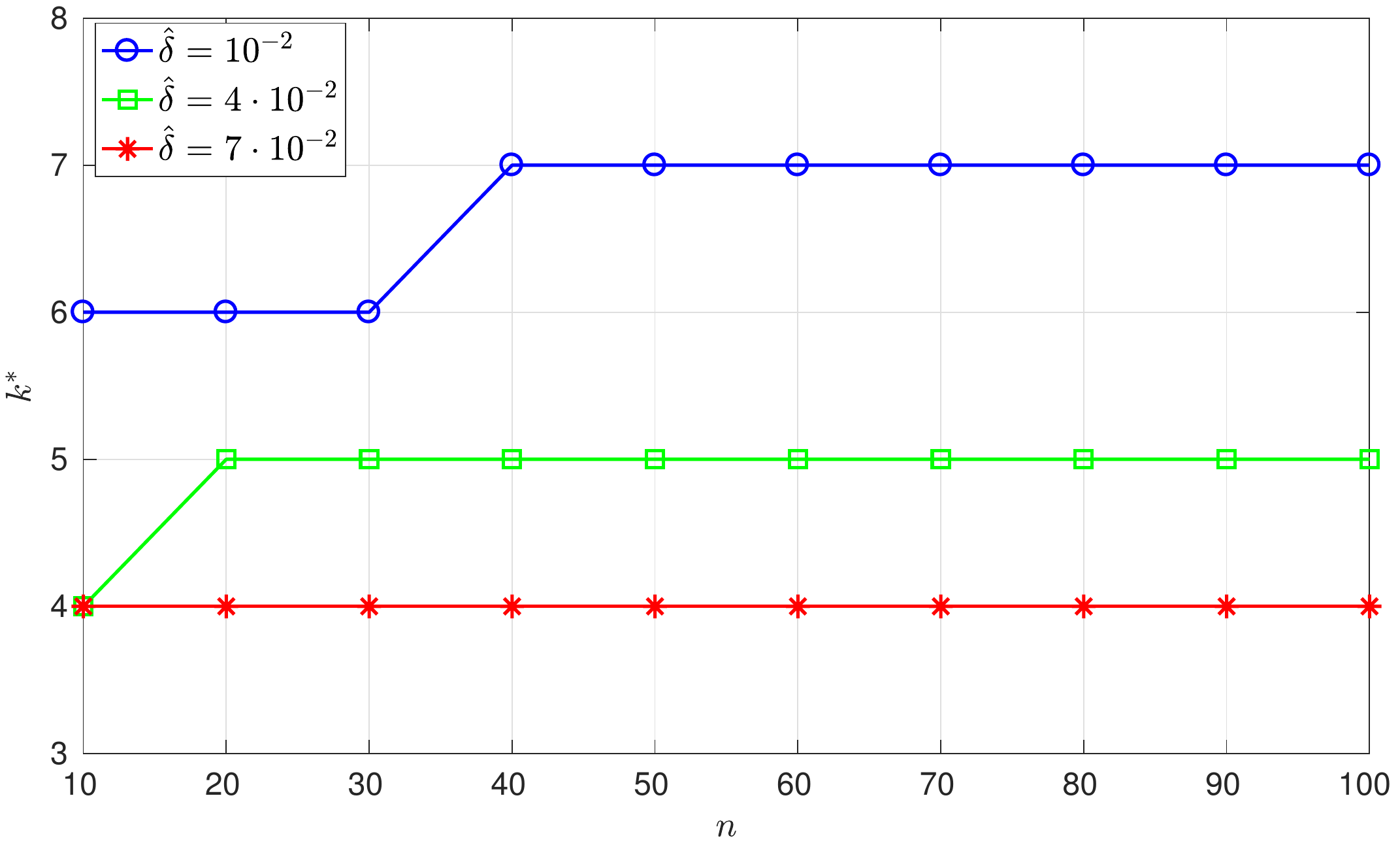}
        \caption{Optimal number of hash functions $k^*$ for different values of $n$ and $\hat{\delta}$.}
        \label{fig:n_vs_k}
    \end{center}
    \vspace{-0pt}
\end{figure}

\begin{figure}[t]
    \begin{center}
        \includegraphics[width=0.9\columnwidth]{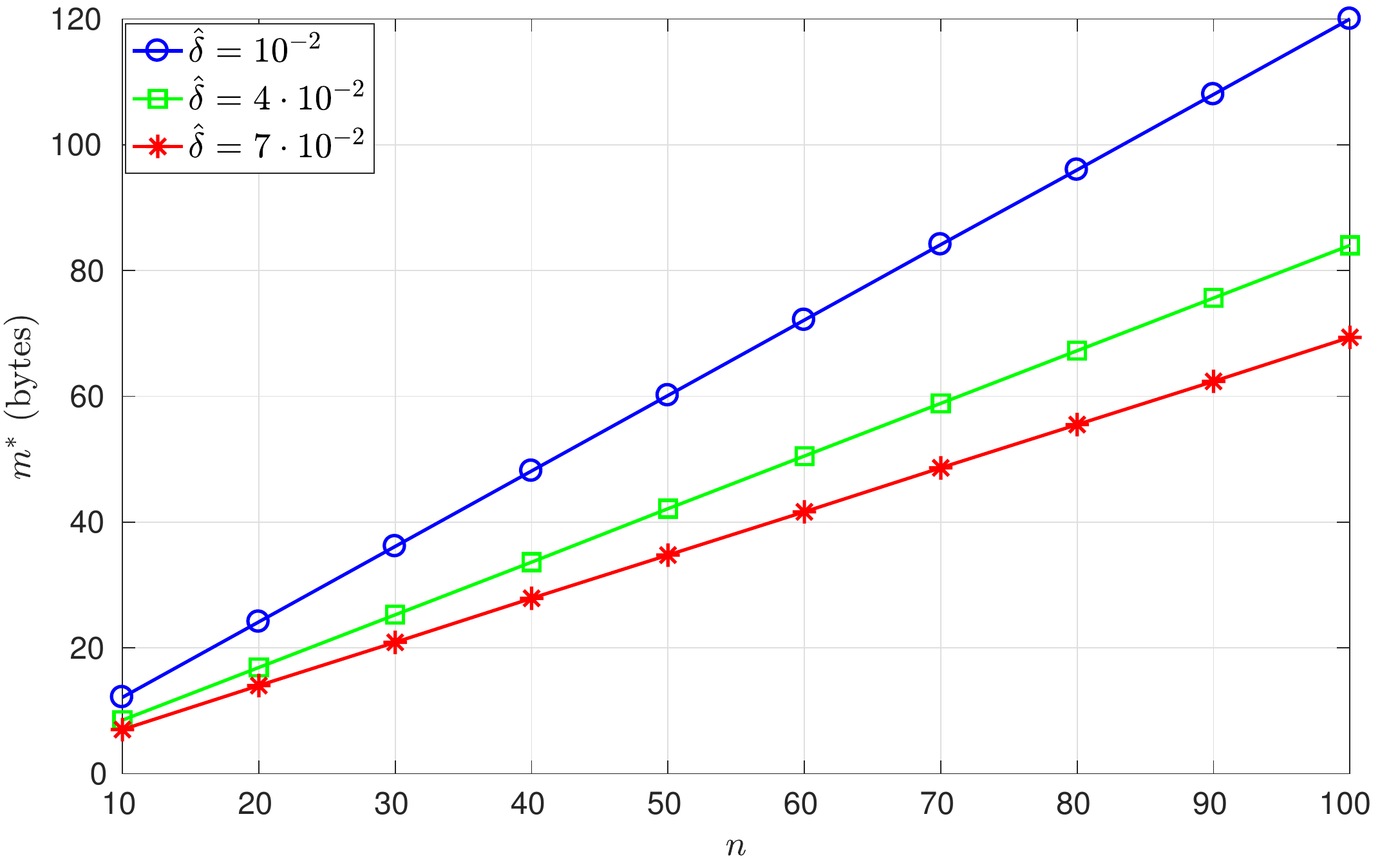}
        \caption{Optimal filter size $m^*$ for different values of $n$ and $\hat{\delta}$.}
        \label{fig:n_vs_m}
    \end{center}
    \vspace{-14pt}
\end{figure}

Fig.~\ref{fig:n_vs_k} shows the optimal number of hash functions $k^*$ for different values of $\hat{\delta}$ and varying input load $n$. As expected, a less strict requirement in terms of false positives leads to a lower and constant $k^*$ as the number of certificates to be compressed becomes higher. On the other hand, by decreasing $\hat{\delta}$ we observe an increment of the optimal number of hash functions, which is necessary to reduce the probability of false positive reports. 
In addition, Fig.~\ref{fig:n_vs_m} shows the optimal filter size $m^*$ (in bits) to meet different values of $\hat{\delta}$ as a function of $n$. We note that $m^*$ linearly increases as the input load becomes higher, thus allowing to keep the false positive rate less than or equal to $\hat{\delta}$. Furthermore, the filter size increase becomes more relevant for $\hat{\delta}=10^{-2}$, as shown by the gap between the blue and green line. This in turn indicates the amount of space required in the filter to prevent multiple hash collisions. 

\subsection{$C^2RL$ v.s. Standard CRL}
\begin{figure}[t]
    \begin{center}
        \includegraphics[width=0.9\columnwidth]{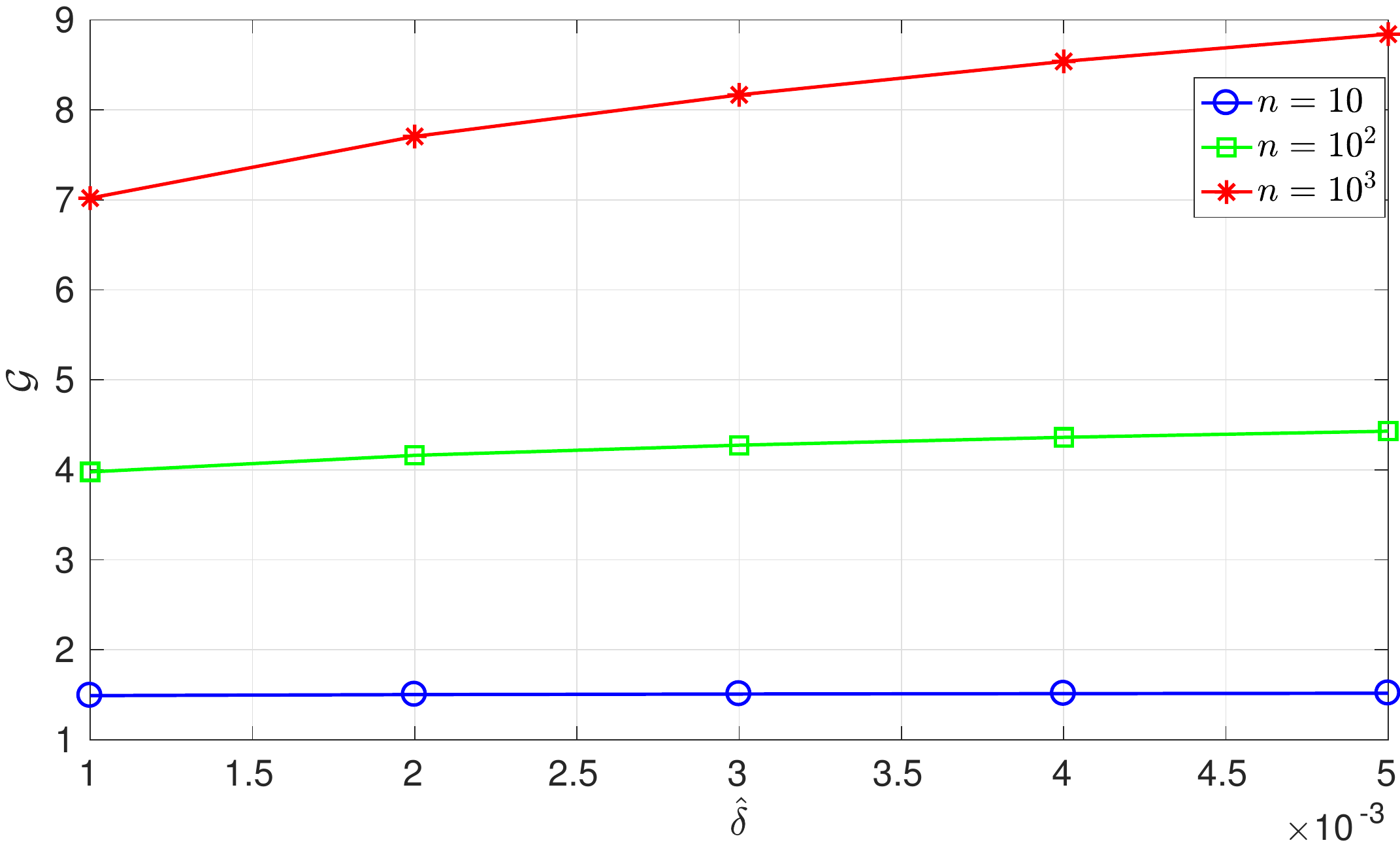}
        \caption{Compression gain as $\hat\delta$ increases and for different values of $n$.}
        \label{fig:Gain}
    \end{center}
    \vspace{-4pt}
\end{figure}

\begin{figure}[t]
    \begin{center}
        \includegraphics[width=0.9\columnwidth]{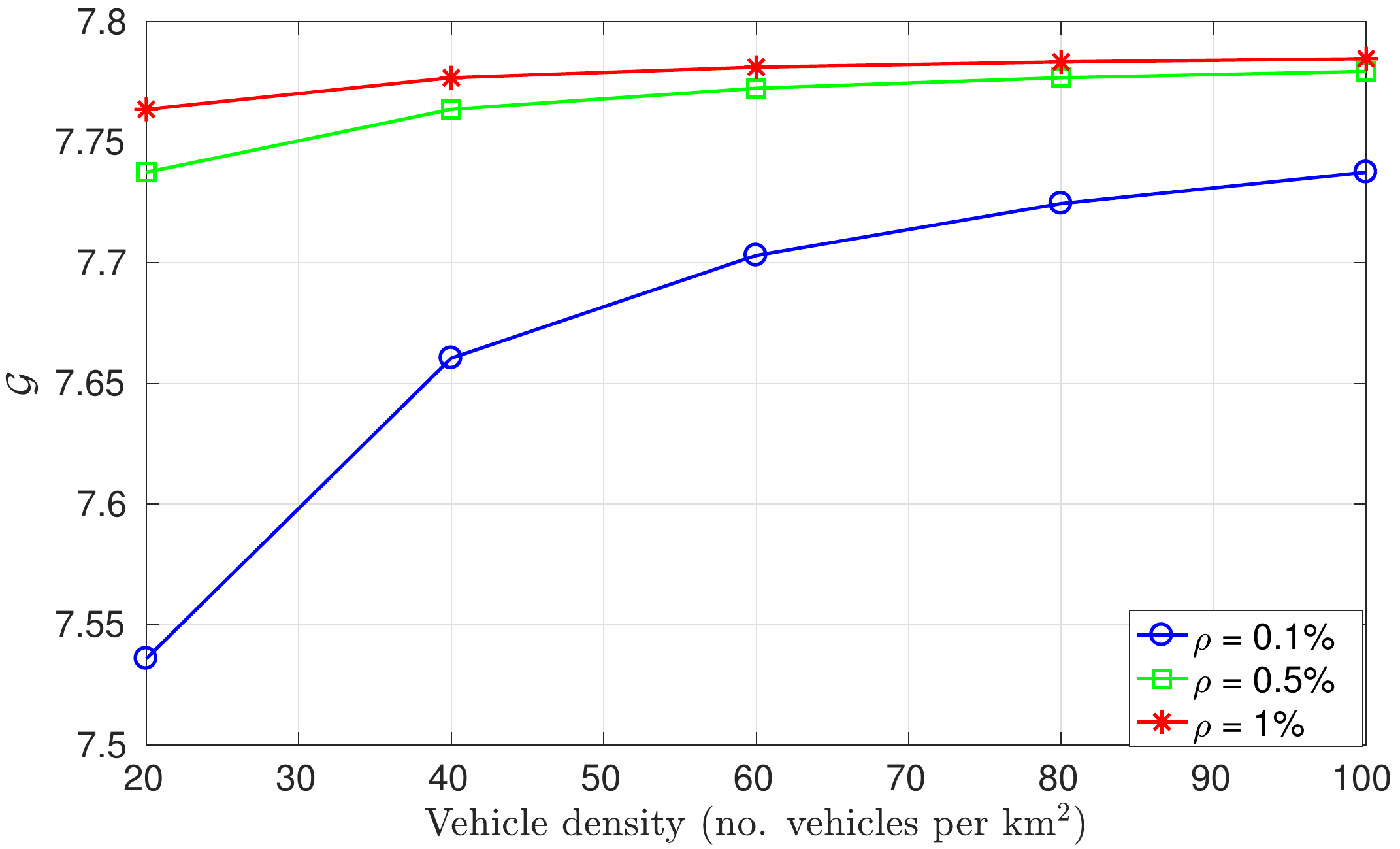}
        \caption{Compression gain as the vehicle density increases and for different values of $\rho$ ($\hat{\delta} = 10^{-3}$).}
        \label{fig:G_vs_density}
    \end{center}
    \vspace{-2pt}
\end{figure}

Fig.~\ref{fig:Gain} shows the compression gain $\mathcal{G}$ as $\hat{\delta}$ increases and for different numbers of certificates revoked $n$. We can note the key benefit provided by the CRL compression, consisting in a significant reduction of the overhead generated by the revocation process. It is also worth pointing out that for low values of $n$ the gain is constant, as the probability of false positives does not significantly influence the choice of the optimal $m$. By contrast, for higher loads, e.g., $n=10^3$, we observe that $\mathcal{G}$ increases from $7$ to $9$, which shows the efficiency of Bloom filters for storing high amounts of certificates as compared with standard CRLs.  
To analyze the performance of the optimized CRL distribution, we also consider an urban vehicular scenario, where vehicles are distributed on an area of 5 km$^{2}$ with different spatial densities, i.e., number of vehicles per km$^2$, managed by a single PCA. 
Furthermore, we assume that each vehicle owns a set of 43800 certificates, which corresponds to the fixed amount of certificates requested by a vehicle driving for two hours and every day of the year~\cite{nowatkowski2010effects}. 
Moreover, we define the revocation rate $\rho$, representing the percentage of vehicles per hour whose certificates must be revoked. 
Fig.~\ref{fig:G_vs_density} shows the average compression gain over an hour, for different values of $\rho$ as the vehicle density increases, and fixed $\hat{\delta}=10^{-3}$. 
We note that for $\rho=0.1\%$, $\mathcal{G}$ considerably increases in the range of densities between 20 and 40, whereas a saturation effect occurs for values of density higher than $60$ vehicles per km$^2$. This is more evident for $\rho$ equal to $0.5\%$ and $1\%$, where the compression gain immediately reaches a saturation level around 7.78. In other words, for higher values of $\rho$ the filter size needs to increase to meet the $\hat{\delta}$ constraint, while  $m^*$ remains low for lower revocation rates. 

\subsection{Large-Scale Urban Scenario}

\begin{figure}[t]
    \begin{center}
        \includegraphics[width=0.9\columnwidth]{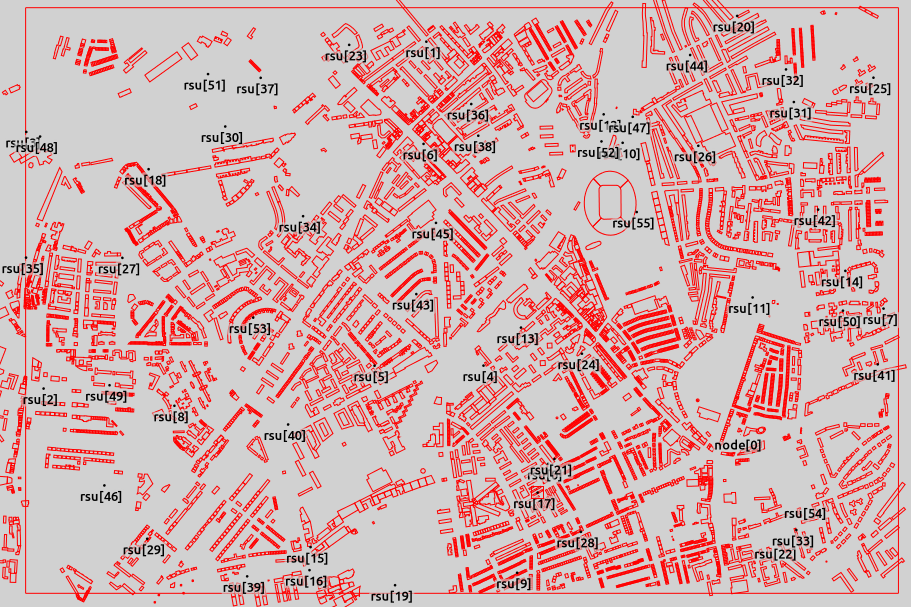}
        \caption{Map of the simulated area in Holloway (London, UK). The figure shows the footprint of the main buildings present in the area and a possible RSU deployment.}
        \label{fig:London_pic}
    \end{center}
    \vspace{-0pt}
\end{figure}

\begin{figure}[t]
    \begin{center}
        \includegraphics[width=0.9\columnwidth]{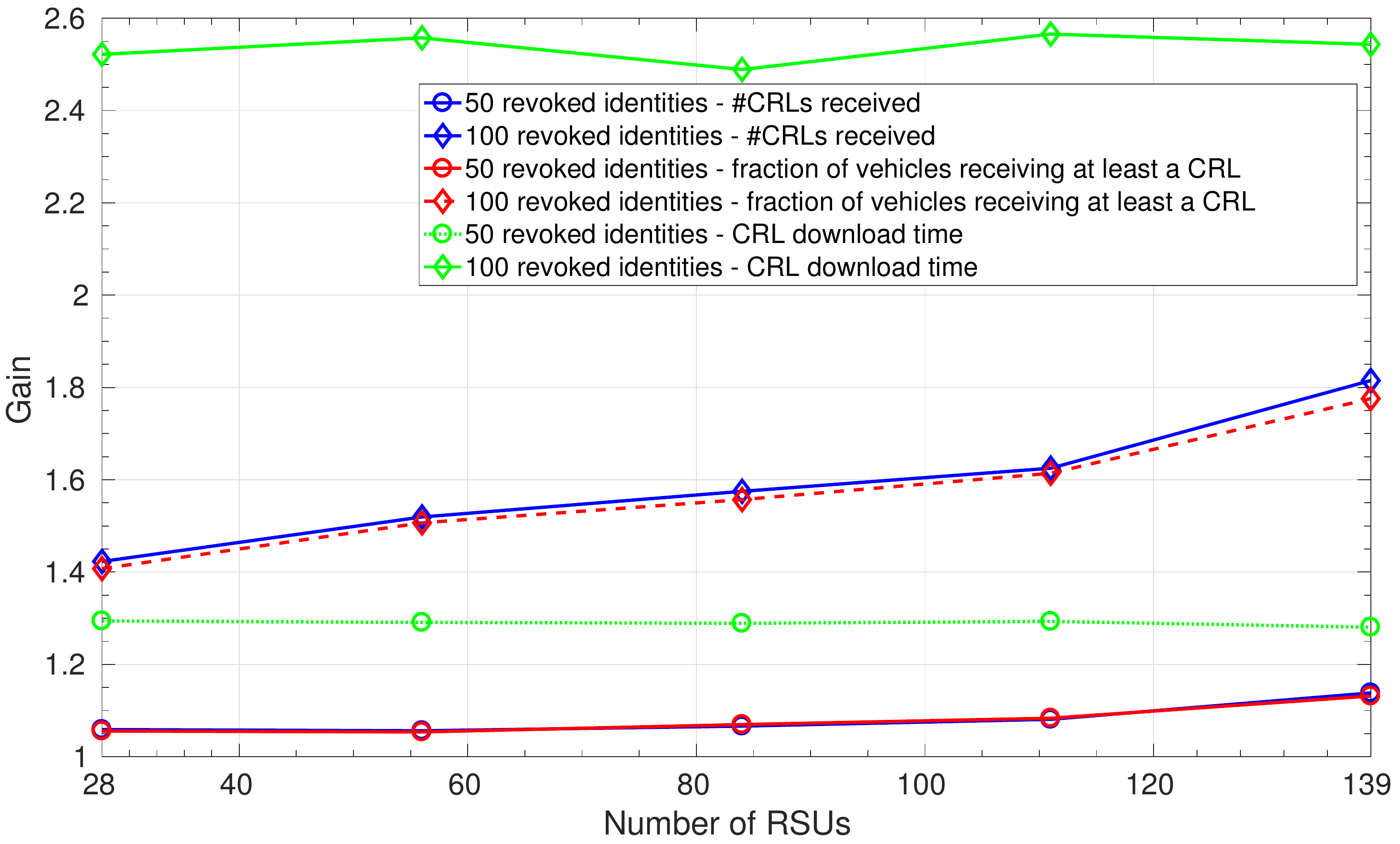}
        \caption{Gains in terms of CRL delivery ratio, percentage of vehicles successfully receiving a CRL and CRL download time.}
        \label{fig:Gains_Veins}
    \end{center}
    \vspace{-0pt}
\end{figure}

To further evaluate the performance gain of the $C^2RL$ scheme over the standard CRL performance, we implemented the aforementioned revocation schemes in a \mbox{OMNet\texttt{++}} network simulator based on the Veins framework~\cite{VEINS}. We considered the urban area of Holloway (London, UK) where a realistic traffic of vehicles has been simulated by means of SUMO~\cite{behrisch2011sumo}. Each vehicle is equipped with a DSRC communication device capable of communicating with RSUs deployed at the side of the road. In order to investigate the impact of different RSU deployments, we placed RSUs uniformly at random in the simulated area preventing the RSUs from being placed within the footprint of a building, as shown in Fig.~\ref{fig:London_pic}. In particular, we considered scenarios ranging from $28$ to $139$ RSUs. Periodically, the CA generates a CRL, which is fragmented by the RSUs, encapsulated in UDP packets and transmitted over the DSRC interface. A list of the relevant simulation parameters is reported in Tab.~\ref{tab:system_parameters}.

Fig.~\ref{fig:Gains_Veins} shows the gains achieved employing the $C^2RL$ scheme over a standard CRL approach in terms of total number of CRLs received by all the vehicles, fraction of vehicles receiving at least one CRL, and average CRL download time as a function of the number of RSUs. The derived performance values have been averaged over multiple instances of RSU deployments. We simulated the cases where $50$ or $100$ vehicles identities per-hours are revoked and each vehicle holds $1000$ pseudonyms. We observe that, in the case of the $C^2RL$ scheme, the Bloom filter parameters $m$ and $k$ have been optimized as in Sec.~\ref{Sec:OptModel}, by referring to $\Hat{\delta} = 10^{-3}$ and a value of $n$ equal to the average number of vehicles present in the simulated area per-second. We note that a higher number of RSUs per-area ensure higher gains for $100$ revoked vehicle identities both in terms of the total number of CRLs received by all the vehicles and the fraction of vehicles receiving at least one CRL - thus ensuring, in the case of $139$ RSUs, gains greater than $1.55$ and $1.6$, respectively. These effects are also detectable for $50$ revoked identities per-hours, yet limited due to the lower traffic load. From Fig.~\ref{fig:Gains_Veins}, we also observe that $C^2RL$ guarantees a shorter CRL download time resulting in an average CRL download time gain greater than $2.5$, for $139$ RSUs and $100$ identities revoked per-hour.


\section{Conclusions}\label{Sec:Conclusions}
In this paper, we proposed an optimized framework to streamline the certificate revocation distribution in a PPKI-based vehicular network. We illustrated the scalability issue resulting from the adoption of a large set of pseudonym certificates per each CAV and discussed the benefits of CRL compression through Bloom filters. 
Significant compression gains can be achieved by adding revoked certificates into a Bloom filter and then disseminating the C$^2$RL in the network. 
We also investigated the impact of different input loads and false positive rates on the optimal choice of $k$ and $m$, and compared our approach with a standard CRL distribution scheme in a realistic large-scale scenario.

\section*{Acknowledgments}
This work is part of the FLOURISH Project, which is supported by Innovate UK under Grant 102582. The authors would like to thank the project collaborators, in particular, Melina Christina, John McCarthy and Ben Miller.

\bibliographystyle{IEEEtran}
\bibliography{biblio.bib}

\end{document}